# U-net super-neural segmentation and similarity calculation to realize vegetation change assessment in satellite imagery[1]

**Chunxue Wu**  WCX@USST.EDU.CN
**Bobo Ju**  XH11407130@OUTLOOK.COM
**Guisong Yang**  GSYANG@USST.EDU.CN
**Hongming Yang**  YHM@USST.EDU.CN
**Jiaying Huang**  SHERRYLUOYE@FOXMAIL.COM
School of Optical-Electrical and Computer Engineering, University of Shanghai for Science and Technology, China, 200093

**Yan Wu**  YANWU8910@GMAIL.COM
Indiana University Bloomington, USA, 47405

**Neal N. Xiong**  XIONGNAIXUE@GMAIL.COM
College of Intelligence and Computing, Tianjin University, China, 300072

**Zhiyong Xu**  ZXU@SUFFOLK.EDU
Math and Computer Science Department, Suffolk University, Boston MA 02108

## Abstract

Vegetation is the natural "linkage" connecting soil, atmosphere and water. It can represent the change of land cover to a certain extent and serve as an "indicator" for global change research. Methods for measuring coverage can be divided into two types: surface measurement and remote sensing. Because vegetation cover has significant spatial and temporal differentiation characteristics, remote sensing has become an important technical means to estimate vegetation coverage. This paper firstly uses U-net to perform remote sensing image semantic segmentation training, then uses the result of semantic segmentation, and then uses the integral progressive method to calculate the forestland change rate, and finally realizes automated valuation of woodland change rate.

## 1. Introduction

In the past few years, the ability of convolutional neural networks to identify objects in many computer vision tasks has surpassed existing human levels. Although convolutional networks(Razavian, et al. 2014) have existed for a long time, their success has been limited due to the size of the training set available for training and the size of the network(Chen, et al. 2018).

Many image segmentation networks are based on improvements in FCNs(Dai, et al. 2016), including U-net(Jaglan, et al. 2019; Ronneberger 2017). U-net consists of two parts, you can see in the Figure 3, the first part, feature extraction, VGG(Iglovikov and Shvets 2018) is similar. The second part is the upsampling part. Since the network structure is U-shaped, it is called a U-net network.

---

[1] The corresponding authors are Chunxue Wu and Zhiyong Xu.

In the U-net feature extraction part, each pooling layers have a scale, including the scale of the original image, there are five scales in total. The U-net upsampling(Burt and Adelson 1987) part, each time upsampling, merges with the same scale feature map corresponding to the feature extraction part, but it is to be cropped before fusion. The fusion here is also spliced.

The U-net network architecture is shown in Fig. 3. It consists of a contracted path (left side) and an extended path (right side). U-net follows the typical architecture of a convolutional network. It consists of two $3\times3$ convolutions, one convolution linear unit (ReLU) per convolution and one $2\times2$ maximum pooling operation, and step 2 for downsampling(Lan, et al. 2015). In each downsampling step, the number of feature channels is doubled. Each step in the expansion path includes upsampling the feature map and then performing a $2\times2$ convolution that halve the number of feature channels, cascade with the response crop feature map from the shrink path, and two $3\times3$ convolutions, Each convolution is followed by a ReLU. Cropping is necessary because each convolution loses boundary pixels. The last layer uses $1\times1$ convolution to map each 64 component feature vectors to the expected number of classes. The network have 23 convolution layers.

In order to achieve seamless tiling of the output segmentation map, it is important to enter the slice size to apply all $2\times2$ maximum pooling operations to even $x$ and $y$ sized layers.

## 2. Related Work

### 2.1 Application of Remapping in Datasets Enhancement

In order to complete the mapping process, it is necessary to obtain some coordinates that are interpolated as non- Remapping(Lecun, et al. 2011; Porterfield 2001; Stearns and Kannappan 1995) is the process of mapping a pixel at a location in one image to a specified location in another image integer pixels because the original image does not have a one-to-one correspondence with the pixel coordinates of the target image. Typically, remapping is used to represent the position $(x,y)$ of each pixel as shown below:

$$g(x,y) = f(h(x,y)) \qquad (1)$$

Here, $g(x,y)$ is the objective function, $f(x,y)$ is the original image, and $h(x,y)$ is the mapping method function acting on $(x,y)$. Let's look at an example. If there is an image $I$, the conditions are remapped:

$$h(x,y) = (I.cols - x, y) \qquad (2)$$

The image will be reversed in the direction of the $x$ axis. Then, the original image and the effect diagram are shown in Fig. 1.

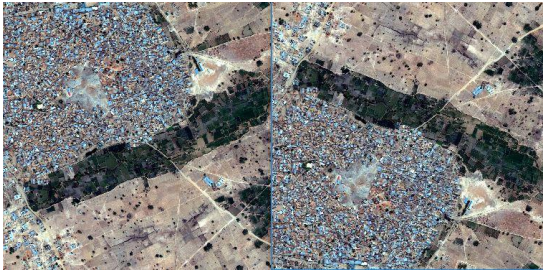

*Figure 1*. 6110_1_2 horizontal flip, with the image center as the origin, perform a flip around the *x* axis for each pixel in the image

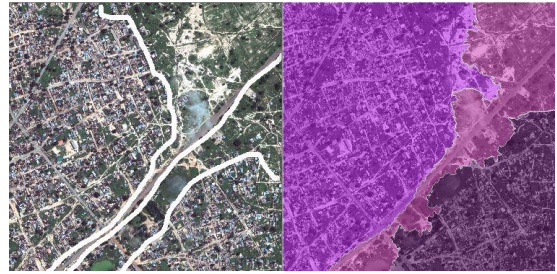

*Figure 2*. 6100_1_3 uses the watershed algorithm to recommend the area of the building area. The left picture shows the edge of the Trees/Crops group through Gaussian blur and edge detection. The right picture shows the segmentation of the building group based on the watershed algorithm.

## 2.2 Regional Recommendations Accelerated Image Segmentation

When segmenting Trees or crops on satellite imagery, the watershed algorithm(Miao, et al. 2016; Zhang, et al. 2016) is used for regional recommendations to avoid time and space consumption of sliding windows in non-building areas and to increase the speed of building segmentation.

In many practical applications, images need to be segmented, but useful information cannot be obtained from the background of images. The watershed algorithm is often very effective in this regard. This algorithm converts the edges in the image into "mountains" and converts the uniform regions into "valleys", which helps to segment the target(Bahadure, et al. 2016).

The watershed calculation process is an iterative labeling process(Huang, et al. 2018). The classical calculation method of watershed is proposed by L. Vincent(Lefranc, et al. 2016). In this algorithm, the watershed calculation is divided into two steps: one is the sorting process and the other is the flooding process. First, the gray level of each pixel is sorted from high to low, and then the flooding process is implemented from low to high. The first-in, first-out (FIFO) structure is used to determine the influence of each local minimum value in the step height and labeling. The watershed transform obtains the image of the collection basin of the input image, and the boundary point between the collection basins is the watershed. Obviously, the watershed represents the maximum point of the input image(Miao and Xiao 2018). The watershed algorithm is shown in Fig. 2.

## 2.3 Similarity measure based on color features

Common similarity measures based on color features mainly include absolute distance, Euclidean distance, histogram intersection method, χ2 distance, reference color table, center distance, and so on(Alsmadi 2017; Bao, et al. 2017; Hafner, et al. 1995).

First, is the feature vector corresponding to the two images, and represents the feature component. The Minkowsky distance(Junior 2006) is defined based on the norm:

$$L(A,B) = \left[\sum_{i=1}^{n}|a_i - b_i|^p\right]^{\frac{1}{p}} \tag{3}$$

(1) If p=1, it is called city-block, which is the absolute distance:

$$L_1(A,B) = \left[\sum_{i=1}^{n}|a_i - b_i|\right] \tag{4}$$

(2) If p=2, it is called Euclidean Distance:

$$L_2(A,B) = \left[\sum_{i=1}^{n}(a_i - b_i)^2\right]^{\frac{1}{2}} \tag{5}$$

## 3. Network Models

The energy function is calculated by combining the pixel-by-pixel soft maximum on the final feature map with the cross entropy loss function(Girshick, et al. 2014; Hariharan, et al. 2015). Softmax is defined as $p_k(x) = \exp(a_k(x)) / \left(\sum_{k}^{K} \exp(a_k(x))\right)$, where $a_k(x)$ represents activation in feature channel $k$ at pixel location $x \in \Omega$, where $\Omega \subset \mathbb{Z}^2$. $K$ is the number of classes, and $p_k(x)$ is the approximate maximum function. That is, $p_k(x) \approx 1$ indicates that $k$ has the largest activation of $a_k(x)$ and $p_k(x) \approx 0$ for all other $k$. Then, the cross entropy penalizes the deviation of $p_{l(x)}(x)$ and $1$ at each position(Ding, et al. 2017).

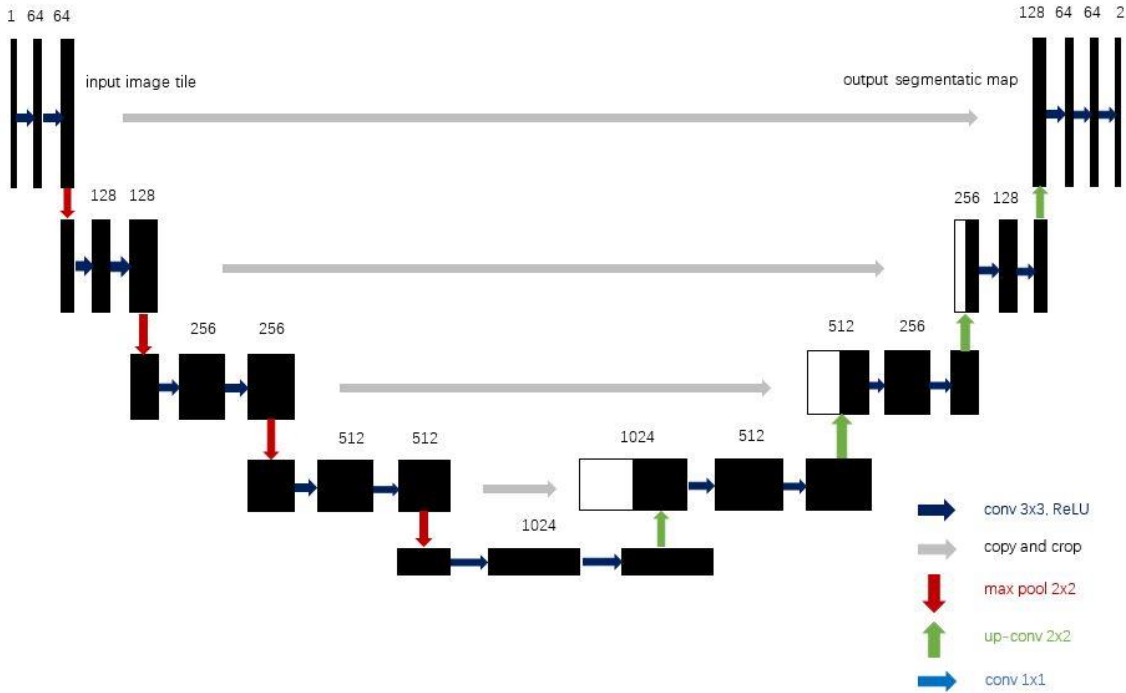

*Figure 3.* The U-net neural network structure consists of a contracted path (left side) and an extended path (right side). It consists of two convolutions, one convolution linear unit (ReLU) and one $2 \times 2$ maximum pooling operation, and step 2 for downsampling. The number of feature channels is doubled in the downsampling step. Each step in the expansion path includes upsampling the feature map and then performing a $2 \times 2$ convolution that halve the number of feature channels, cascade with the response crop feature map from the shrink path, and two $3 \times 3$ convolutions, Each convolution is followed by a ReLU, and the last layer uses $1 \times 1$ convolution to map each 64 component feature vectors to the expected number of classes.

$$E = \sum_{x \in \Omega} w(x) \log\left(p_{l(x)}(x)\right) \tag{6}$$

$l : \Omega \to \{1, ..., K\}$ is the real label for each pixel, and $w : \Omega \to \mathbb{R}$ is the weight map be introduced to

make some pixels more important in training. The weight map for each satellite image real-time segmentation is pre-computed to compensate for different pixel frequencies of a certain category in the training datasets and forces the network to learn small separation boundaries that are introduced between touch cells (see Figures 3c and d)).

Morphological operations(Bai, et al. 2017) are used to calculate the separation boundary. Then calculate the weight map as

$$w(x) = w_c(x) + w_0 \exp\left(-\frac{(d_1(x) + d_2(x))^2}{2\sigma^2}\right) \tag{7}$$

Where $w_c : \Omega \to \mathbb{R}$ is the weight map of the balanced class frequency, $d_1 : \Omega \to \mathbb{R}$ represents the distance to the nearest cell boundary, and $d_2 : \Omega \to \mathbb{R}$ represents the distance to the second nearest Trees/Crops boundary. In the experiment, the $w_0 = 10$ and $\sigma \approx 5$ pixels are set(Abdalla and Nagy 2018; Bai and Liu 2017).

In a deep network with many convolutional layers and different paths through the network, good initialization of the weights is very important(Xue, et al. 2018). Otherwise, some parts of the network may be activated too much, while others will not contribute. Ideally, the initial weights should be adjusted so that each feature map in the network has an approximate unit variance(Xi, et al. 2018). For networks with this architecture (alternating convolution and ReLU layers), this can be achieved by plotting initial weights from a Gaussian distribution with standard deviation $\sqrt{2/N}$, where $N$ represents the number of incoming nodes for a neuron. For example, for the $3 \times 3$ convolution in the previous layer and $64$ feature channels $N = 9 \cdot 64 = 576$.

When only a small number of samples are available, data enhancement is critical to the invariance and robustness properties required to train the network. In the case of segmentation of remote sensing images, horizontal flipping, vertical flipping, random rotation and color blurring are applied randomly to increase the number of images in the batch. Although there are many methods such as undersampling and adjusting the weight of the datasets when dealing with the unbalanced problem of the datasets, the oversampling method above is the most effective and simplest method(Krizhevsky, et al. 2012; Lecun, et al. 2011; Porterfield 2001).

The method of the whole study is shown in figure 4.

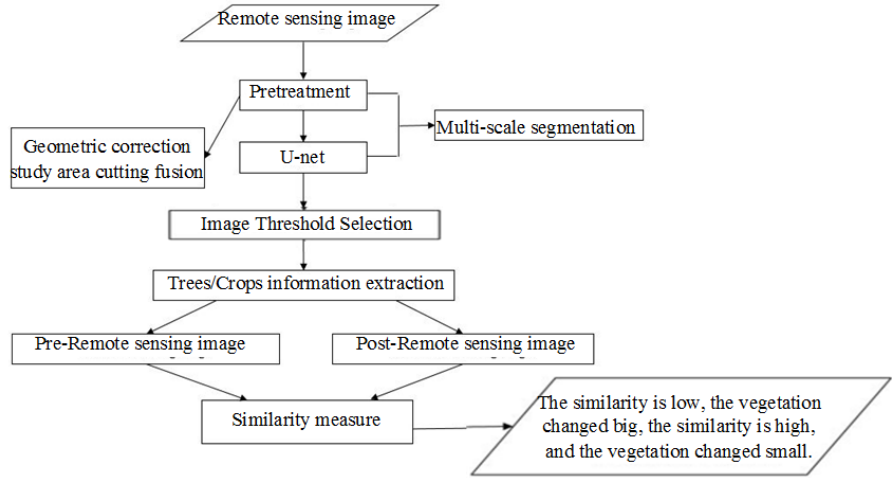

*Figure 4.* U-net super-neural segmentation and similarity calculation to realize vegetation change assessment in satellite imagery.

*Table 1*. Datasets Contains Content Overview.

| TYPES | BANDS | PIXEL RESOLUTION | CHANNELS | SCALE |
|---|---|---|---|---|
| GRAYSCALE | PANCHROMATIC BAND | 0.31m | 1 | 3348x3392 |
| 3 CHANNELS | RGB | 0.31m | 3 | 3348x3392 |
| 16 CHANNELS | MULTISPECTRAL | 1.24m | 8 | 837x848 |
| 16 CHANNELS | NEAR INFRARED | 7.5m | 8 | 134x136 |

## 4. Analysis of the Experiment and Results

### 4.1 Datasets

The datasets used in this trial was provided by the UK National Defense Science and Technology Laboratory (Fig. 5 shows a partial 3-channel satellite imagery of the dataset). The training set contains 25 high-resolution satellite images of 1 square kilometer area. There are three versions of the training set image to choose from, namely grayscale, 3-channel RGB color map and 16-channel image. See Table 1 for details.

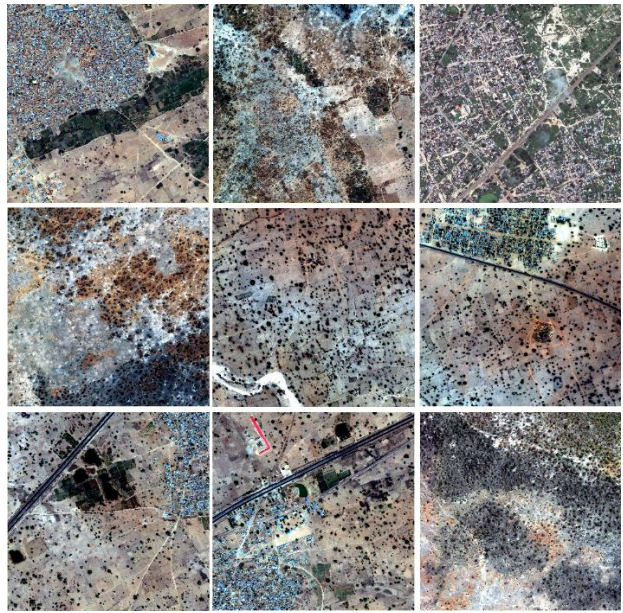

*Figure 5*. It is a satellite image of all the original datasets cited in the text (the first line from left to right image ID numbers are: 6110_1_2, 6010_1_2, 6100_1_3, the second line from left to right image ID numbers are: 6010_4_2, 6060_2_3, 6120_2_0, the second line from left to right image ID numbers are: 6110_4_0. 6110_3_1, 6040_2_2).

By adjusting and aligning the 16-channel image to match the 3-channel RGB image, a calibration operation is necessary to eliminate the difference between the channels. Finally, the training focuses on three versions of the image and integrates it into a 20-channel input image.

**4.2 Experimental Methods and Processes**

The preprocessed image is left unchanged according to the actual category of the image, or the image and its corresponding label are collectively adjusted to a square image of $1920 \times 1920$ resolution. During the training, image blocks of size $144 \times 144$ are randomly collected from different images, and half of the image blocks always contain some positive pixels, ie, classified target objects. Each network model has approximately 1.7 million parameters and a batch size of 60. This network training does not use existing models for fine-tuning.

In the prediction process, the sliding window method is used. The window size is fixed to $144 \times 144$ and the sliding step is 32. The object of the image block boundary can only be partially displayed when there is no surrounding environment. The above operation can eliminate
prediction of the weak effect of the image block boundary. To further improve the quality of the prediction, the flipped and rotated versions of the input image, as well as the network models trained at different sizes, are averaged. All in all, a very smooth output is obtained.

The input image and its corresponding segmentation map are used to train the network, using TensorFlow's stochastic gradient descent implementation. Since it is a convolution without filling, the output image is smaller than the constant boundary width of the input. To minimize overhead and maximize memory utilization, large batches of large input slices tend to be used to reduce the batch size to a single image. Therefore, high momentum (0.99) is used to utilize a large number of previous training samples to determine updates in the current optimization step.

Since the image similarity measure has strong subjectivity, it is not easy to objectively evaluate the performance of an image similarity measure algorithm. When performing image similarity metrics, there are one or more characterization methods and similar metrics, which require a comprehensive evaluation of the similarity measures of different image features or feature combinations and different similarity measures. Here, The binary image change of woodland in remote sensing image is extracted (Same geographic location), and then using Minkowsky distance to measure image similarity. See Appendix A for specific methods.

**4.3 Experimental Results and Analysis**

In the end, the Trees model training accuracy in this test reached 63%, and the test accuracy reached 69%. the Trees model training accuracy in this test reached 83%, and the test accuracy reached 88%. The analysis of the training results in the TensorBoard is shown in Figs 6, 7, 8 and 9. By comparing these two numbers, it can be seen that the trained model and the pre-processed enhanced datasets are matched, and neither overfitting nor underfitting occurs during training. When the training reaches 9000/16000 cycles, the model is close to convergence, and its test accuracy is close to the training accuracy, which reflects that the model is good.

According to the previous algorithm model, in the end, two-hole satellite image maps with dense Trees were tested (as shown in Figs. 10 and 11), two-hole satellite image maps with dense Trees were tested (as shown in Figs. 12 and 13). Each satellite image shows the geomorphology of different areas of 1 square kilometer. Test results The top two images show the area of the Trees/Crops marked by the real label, and the two figures in the lower part show the model predictions. The result of the labeling of the Trees/Crops on the left sub-picture in the image is superimposed on the real satellite image on the right side by adding a layer. As shown in Tables 2 and 3, from the comparison of the actual results and the predicted results of the three test images, it can concluded that although the model can not accurately mark all the Trees/Crops in the satellite image during the process of predicting the satellite image, but the model can be compared with real artificially labeled Trees/Crops! In the process of segmentation, the model can accurately distinguish the complex environment around the Trees/Crops (including some other artificial mixed buildings), and filter out some noise and non-constructive buildings which even make human confused (Prove that the threshold set during image segmentation is reasonable). In actual use, the model can predict Trees/Crops in other satellite imagery very well. This is due to the tremendous achievements that neural networks have made in image recognition in recent years, especially success of U-net, used for hyper-neural segmentation network models. It also proves that it is very appropriate to pre-process the dataset in the early stage. Otherwise, the satellite image datasets is too small, and the overfitting or underfitting can occurs during the training process, which can lead to the whole model being very bad or even completely impossible predict the area of the Trees/Crops in the new image. U-net theoretically takes about 87 seconds to segment the pixel image, but in fact the watershed algorithm is used to pre-exclude some non-building areas. Many $144 \times 144$ disassembled for each $3396 \times 3349$ image. Identification, and re-splicing many recognition results into the original image, memory loading of the model and satellite imagery, these are extremely time-consuming. The time actually used for segmentation is much longer than 87 seconds (Ding, et al. 2017). Perhaps the experimental equipment used is not good enough, and training with a higher performance server can achieve better recognition speed.

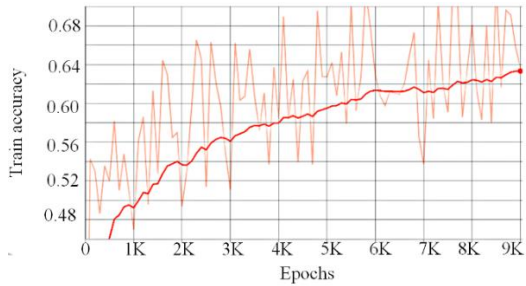

*Figure 6.* After 9000 iterations of the model, the training accuracy is 63%. From the figure, it can be seen that the training accuracy of the image is growing fast from 0 to 6000 iterations. The learning rate is reduced by 10 times every 4000 cycles, and the training accuracy of the model is very close to 63% after 8000 cycles of iteration.

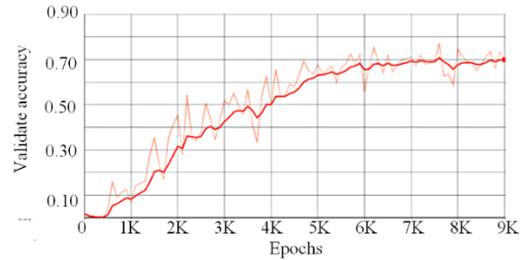

*Figure 7.* After 9000 iterations of the model, the training accuracy is 69%. From the figure, it can be seen that the training accuracy of the image is growing fast from 0 to 6000 iterations. The learning rate is reduced by 10 times every 4000 cycles, and the training accuracy of the model is very close to 69% after 8000 cycles of iteration.

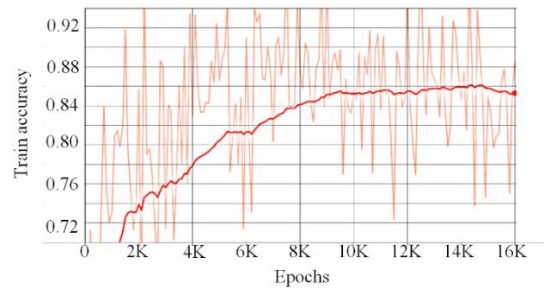

*Figure 8.* After 16000 iterations of the model, the training accuracy is 83%. From the figure, it can be seen that the training accuracy of the image is growing fast from 0 to 10000 iterations. The learning rate is reduced by 10 times every 4000 cycles, and the training accuracy of the model is very close to 83% after 8000 cycles of iteration.

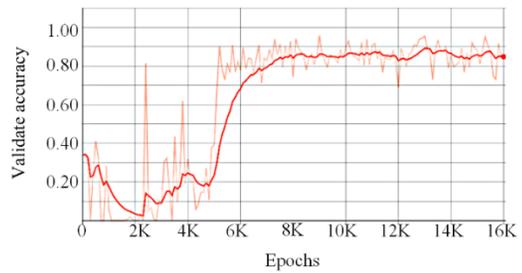

*Figure 9.* After 9000 iterations of the model, the training accuracy is 88%. From the figure, it can be seen that the training accuracy of the image is growing fast from 0 to 1500 iterations. The learning rate is reduced by 10 times every 4000 cycles, and the training accuracy of the model is very close to 88% after 8000 cycles of iteration.

Table 2. Comparison of the Number of Real Trees and the Number of Trees Predicted by the Model.

| SATELLITE IMAGE ID | ACTUAL NUMBER OF TREES | FORECASTED NUMBER OF TREES |
|---|---|---|
| 6010_4_2 | 2262 | 2177 |
| 6060_2_3 | 1613 | 1574 |

Table 3. Comparison of the Number of Real Crops and the Number of Crops Predicted by the Model.

| SATELLITE IMAGE ID | ACTUAL NUMBER OF CROPS | FORECASTED NUMBER OF CROPS |
|---|---|---|
| 6110_4_0 | 29 | 29 |
| 6110_3_1 | 37 | 38 |

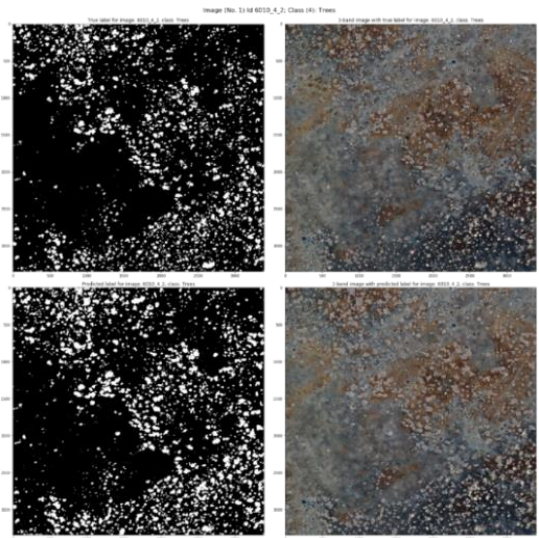

Figure 10. Group 1: 1-km 6010_4_2 TIFF satellite imagery test results. In the four subgraphs (the first layer is 6010_4_2 (a), 6010_4_2 (b) from left to right; the second layer is 6010_4_2 (c), 6010_4_2 (d) from top left to right.) (a), (b) are the results of the labeling of the real labels in the datasets, the lower half (c), (d) are the model prediction results.

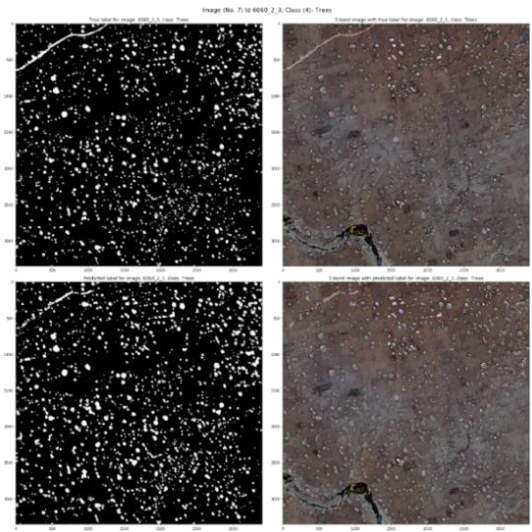

Figure 11. Group 1: 1-km 6060_2_3 TIFF satellite imagery test results. In the four subgraphs (the first layer is 6060_2_3 (a), 6060_2_3 (b) from left to right; the second layer is 6060_2_3 (c), 6060_2_3 (d) from top left to right.) (a), (b) are the results of the labeling of the real labels in the datasets, the lower half (c), (d) are the model prediction results.

THE SEVENTH ANNUAL CONFERENCE ON ADVANCES IN COGNITIVE SYSTEMS

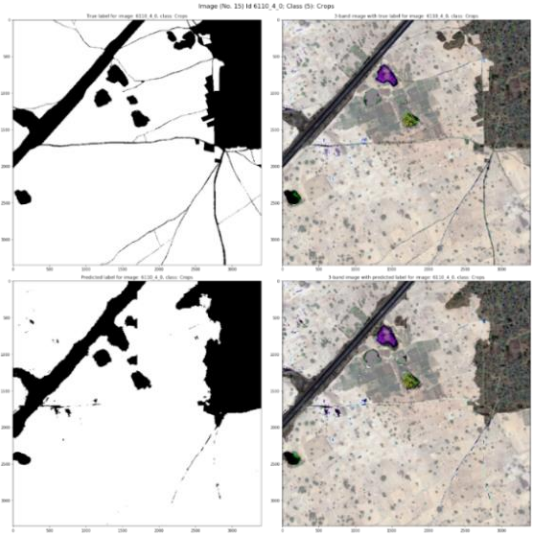 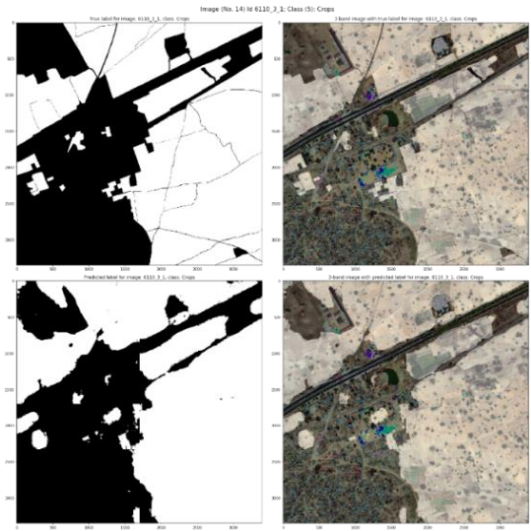

*Figure 12.* Group 1: 1-km 6110_4_0 TIFF satellite imagery test results. In the four subgraphs (the first layer is 6110_4_0 (a), 6110_4_0(b) from left to right; the second layer is 6110_4_0(c), 6110_4_0(d) from top left to right.) (a), (b) are the results of the labeling of the real labels in the datasets, the lower half (c), (d) are the model prediction results.

*Figure 13.* Group 1: 1-km 6110_3_1 TIFF satellite imagery test results. In the four subgraphs (the first layer is 6110_3_1 (a), 6110_3_1 (b) from left to right; the second layer is 6110_3_1 (c), 6110_3_1 (d) from top left to right.) (a), (b) are the results of the labeling of the real labels in the datasets, the lower half (c), (d) are the model prediction results.

## 5. Conclusion and Future Work

In the field of remote sensing applications, vegetation index is an important source of information reflecting surface vegetation information, and has been widely used to qualitatively and quantitatively evaluate vegetation cover and its growth vitality. This paper proves that the evaluation of vegetation coverage by neural network can achieve good results through experimental analysis. The U-Net method mentioned in the previous sections will bring significant IoU gain if it is replaced by its variant U-Net++. In recent research, we intend to find a kind of Dilated Stacked U-Net++s architecture to improve the segmentation accuracy and fine-grained details performance of forest land in satellite imagery.

## Acknowledgement

The authors would like to appreciate all anonymous reviewers for their insightful comments and constructive suggestions to polish this paper in high quality. This research was supported by the National Key Research and Development Program of China (No.2018YFC0810204), National Natural Science Foundation of China (No. 61872242, 61502220), Shanghai Science and Technology Innovation Action Plan Project (17511107203,16111107502) and Shanghai key lab of modern optical system.

**Appendix A** (Calculating rate of regional vegetation change using progressive method of integration).

This experiment uses semantic segmentation to extract vegetation coverage areas from satellite images of different years in the same region and generate a binary map (as shown in Figures 1 and 2). Split the binary image into tile regions. Then use the color similarity to calculate the number of tiles with white vegetation coverage area, and obtain the vegetation coverage area. The vegetation coverage calculation method is as follows:

$$T = \sum_{0}^{n} \int_{a(i)}^{b(i)} [f_i(x) - g_i(x)] d(x) \quad (i = 0,\ldots,n) \tag{8}$$

$T_q$ is the original vegetation area, and $T_h$ is the changed vegetation area in the same area. The vegetation cover change rate is calculated as follows:

$$\mu = \frac{T_h - T_q}{T_q} \tag{9}$$

If $\mu$ is positive, it means that the vegetation coverage in this area is increasing, and the larger the absolute value of $\mu$, the faster the vegetation coverage grows. If $\mu$ is negative, it means that the vegetation cover in the area is degraded, and the larger the absolute value of $\mu$, the faster the vegetation cover degradation. If $\mu$ is 0, there is no change in vegetation in the area.

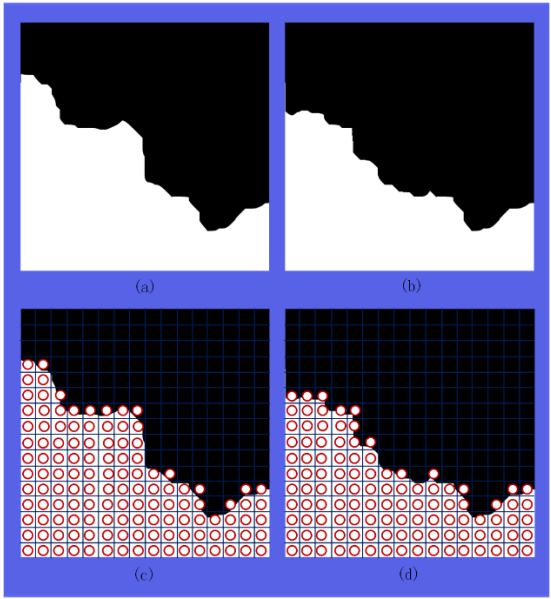

*Figure 14.* Degradation of vegetation edges in a region. (a) and (c) are the vegetation cover of the area before degeneration. (b) and (d) are the vegetation cover in the area after degradation. (c) is a progressive representation of the integral of (a). (d) is a progressive representation of the integral of (b). Before degradation, (c) corresponding to (a), there are 126 tile areas with vegetation; after degradation, (b) corresponds to (d), there are 110 tile areas with vegetation. The vegetation degradation rate in the area is 12.7%.

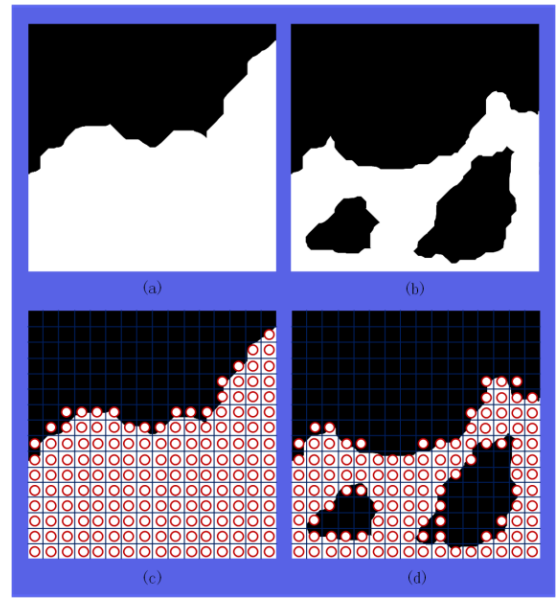

*Figure 15.* Inside degradation of vegetation in a region. (a) and (c) are the vegetation cover of the area before degeneration. (b) and (d) are the vegetation cover in the area after degradation. (c) is a progressive representation of the integral of (a). (d) is a progressive representation of the integral of (b). Before degradation, (c) corresponding to (a), there are 168 tile areas with vegetation; after degradation, (b) corresponds to (d), there are 121 tile areas with vegetation. The vegetation degradation rate in the area is 28.0%.